\documentclass[11pt]{article}
\pdfoutput=1
\usepackage{jheppub}
\usepackage{amsmath,amssymb,amsthm,bm,bigdelim,multirow}
\usepackage{color}
\usepackage{booktabs}
\usepackage{hyperref}

\newcommand{\kslash}{k\kern-1ex /}
\newcommand{\pslash}{p\kern-1ex /}
\newcommand{\qslash}{q\kern-1ex /}
\newcommand{\lslash}{l\kern-1ex /}
\newcommand{\sslash}{s\kern-1ex /}
\newcommand{\Dslash}{D\kern-1.2ex /}

\newcommand{\beqa}{\begin{eqnarray}}
\newcommand{\eeqa}{\end{eqnarray}}

\newcommand{\bd}{\begin{description}}
\newcommand{\ed}{\end{description}}
\newcommand{\la}{\langle}
\newcommand{\ra}{\rangle}
\newcommand{\ben}{\begin{eqnarray}}
\newcommand{\een}{\end{eqnarray}}

\def\lsim{\raise0.3ex\hbox{$<$\kern-0.75em\raise-1.1ex\hbox{$\sim$}}}
\def\gsim{\raise0.3ex\hbox{$>$\kern-0.75em\raise-1.1ex\hbox{$\sim$}}}
\def\simgt{\rlap{\lower 3.5 pt\hbox{$\mathchar \sim$}}\raise 2.0pt \hbox {$>$}}
\def\simlt{\rlap{\lower 3.5 pt\hbox{$\mathchar \sim$}}\raise 2.0pt \hbox {$<$}}

\begin{document}
  \title{Tensor renormalization group study of two-dimensional U(1) lattice gauge theory with a $\theta$ term}

  \author[a]{Yoshinobu Kuramashi}
%  \affiliation{Faculty of Pure and Applied Sciences, University of Tsukuba, Tsukuba, Ibaraki 305-8571, Japan}
%  \affiliation{Center for Computational Sciences, University of Tsukuba, Tsukuba, Ibaraki  305-8577, Japan}
%  \affiliation{RIKEN Advanced Institute for Computational Science, Kobe, Hyogo 650-0047, Japan}
 
  \author[a]{Yusuke Yoshimura}
%  \affiliation{RIKEN Advanced Institute for Computational Science, Kobe, Hyogo 650-0047, Japan}
%  \affiliation{Graduate School of System Informatics, Department of Computational Sciences, Kobe University, Kobe, Hyogo 657-8501, Japan}
  \affiliation[a]{Center for Computational Sciences, University of Tsukuba, Tsukuba, Ibaraki
    305-8577, Japan}

%\begin{abstract}
%We make an analysis of the two-dimensional U(1) lattice gauge theory with a $\theta$ term by using the tensor renormalization group. Our numerical result for the free energy shows good consistency with the exact one at finite coupling constant. The topological charge density generates a finite gap at  $\theta=\pi$ toward the thermodynamic limit. In addition finite size scaling analysis of the topological susceptibility up to $V=L\times L=1024\times 1024$ allows us to determine the phase transition at $\theta=\pi$ is the first order. 
%\end{abstract}
%\pacs{05.10.Cc, 11.15.Ha}
\date{\today}

\preprint{UTHEP-738, UTCCS-P-125}

\maketitle

\section{Introduction}
\label{sec:intro}

It has been argued that pure gauge theories with a $\theta$ term contain intriguing nonperturbative aspects. Possible phase transition in the two-dimensional (2D) pure U($N$) gauge theory was investigated  at $\theta=0$ in the large $N$ limit by Gross and Witten thirty years ago \cite{un_2d} and Seiberg discussed that it has a phase transition at $\theta=\pi$ in the strong coupling limit \cite{cpn_b0}. Later Witten showed that the four-dimensional (4D) pure Yang-Mills theory yields the spontaneous CP violation at $\theta=\pi$ in the large $N$ limit \cite{sun_4d_n}. Recently this non-trivial phenomena was also predicted based on the argument of the anomaly matching between the CP symmetry and the center symmetry \cite{sun_4d_anom}. Up to now, unfortunately, the numerical study with the lattice formulation has not been an efficient tool to investigate these nonperturbative phenomena. The reason is that the lattice numerical methods are based on the Monte Carlo algorithm so that they suffer from the sign problem caused by the introduction of the $\theta$ term.   

In 2007 the tensor renormalization group (TRG) was proposed by Levin and Nave to study 2D classical spin models \cite{trg}. They pointed out that the TRG method does not suffer from the sign problem in principle. This is a fascinating feature to attract the attention of the elementary particle physicists, who have been struggling with the sign problem to investigate the finite density QCD, the strong CP problem, the lattice supersymmetry and so on. In past several years exploratory numerical studies were performed by applying the TRG method to the quantum field theories in the path-integral formalism \cite{phi4,schwinger,schwinger-theta,schwinger-phase,njl-fd,cpn,n1-wz,gtrg_3d,gtrg_gf,u1higgs,phi4_quad,z2gauge_3d,schwinger-g,cphi4}. 
%A difficulty exists in developing efficient algorithms for the scalar, fermion and gauge theories. For the scalar part we have employed the Gauss-Hermite quadrature to discretize the continuous degree of freedom of the scalar field \cite{n1-wz}. For the fermionic system we have developed a Grassmann version of the tensor renormalization group to treat the fermion field in the path-integral formalism \cite{schwinger,schwinger-theta,schwinger-phase,n1-wz}. Armored with these techniques
The authors and their collaborators have  confirmed that the TRG method is free from the sign problem by successfully demonstrating the phase structure predicted by Coleman \cite{schwinger-theta_anal} for the one-flavor Schwinger model with the $\theta$ term employing the Wilson fermion formulation \cite{schwinger-theta}\footnote{See Ref.~\cite{funcke} for recent studies of the Schwinger model with the $\theta$ term in the Hamiltonian formalism.} and the Bose condensation accompanied with the Silver Blaze phenomena in the 2D complex scalar $\phi^4$ theory at the finite density \cite{cphi4}. 

In this article we apply the TRG method to the 2D pure U(1) lattice gauge theory with a $\theta$ term. Since this is the simplest pure lattice gauge theory with a $\theta$ term and the analytical result for the partition function is already known \cite{u1_anal}, it is a good test case for the TRG method to check the feasibility to investigate the nonperturbative properties of the lattice gauge theories with a $\theta$ term. In the previous studies of Schwinger model with and without the $\theta$ term \cite{schwinger,schwinger-theta,schwinger-phase}, we employed the character expansion method to construct the tensor network representation following the proposal in Ref.~\cite{tn-rep}. In this work, however, we use the Gauss quadrature method with some improvement to discretize the phase in the U(1) link variable. This is motivated by the success of the Gauss quadrature method to discretize the continuous degree of freedom in the TRG studies of the scalar field theories \cite{phi4_quad,cphi4}.

This paper is organized as follows. In Sec.~\ref{sec:method} we explain the TRG method with the use of the Gauss quadrature to calculate the partition function of the 2D pure U(1) gauge theory. Numerical results for the phase transition at $\theta=\pi$ are presented in Sec.~\ref{sec:analysis}, where our results are compared with the exact ones which are analytically obtained.  Section~\ref{sec:summary} is devoted to summary and outlook.

\section{Tensor renormalization group algorithm}
\label{sec:method}

\subsection{2D pure U(1) lattice gauge theory with a $\theta$ term}
The Euclidean action of the two-dimensional pure U(1) lattice gauge theory with a $\theta$ term is defined by
\begin{gather}
	S= -\beta \sum_x \cos p_x -i\theta Q, \\
	p_x= \varphi_{x,1}+\varphi_{x+\hat 1,2}-\varphi_{x+\hat 2,1}-\varphi_{x,2}, \\
	Q= \frac{1}{2\pi} \sum_x q_x, \quad q_x=p_x \bmod 2\pi,
\end{gather}
where $\varphi_{x,\mu}\in [-\pi,\pi]$ is the phase of U(1) link variable at site $x$ in $\mu$ direction.
The range of $q_x$ is $[-\pi,\pi]$ and it can be expressed as follows by introducing an integer $n_x$:
\begin{align}
	q_x =p_x +2\pi n_x, \quad n_x\in \{-2,-1,0,1,2\}.
\end{align}
For the periodic boundary condition, the topological charge $Q$ becomes an integer:
\begin{align}
	Q =\sum_x \left( \frac{p_x}{2\pi} +n_x \right) =\sum_x n_x
\end{align}

The tensor may be given with continuous indices,
\begin{align}
	\mathcal T(\varphi_{x,1},\varphi_{x+\hat 1,2},\varphi_{x+\hat 2,1},\varphi_{x,2})
	=\exp\left( \beta\cos p_x +i\frac{\theta}{2\pi}q_x \right).
\end{align}
The partition function is represented as
\begin{align}
	Z =\left( \prod_{x,\mu}\int_{-\pi}^\pi \frac{d\varphi_{x,\mu}}{2\pi} \right)
	\prod_x \mathcal T(\varphi_{x,1},\varphi_{x+\hat 1,2},\varphi_{x+\hat 2,1},\varphi_{x,2}).
	\label{eq:PF}
\end{align}

\subsection{Gauss-Legendre quadrature method}
In order to obtain a finite dimensional tensor network,
we discretize all the integrals in Eq.~(\ref{eq:PF}) using a numerical quadrature.
In general, an integral of a function $f(\varphi)$ can be evaluated by
\begin{align}
	\int d\varphi f(\varphi)
	\approx \sum_{\alpha=1}^K w_\alpha f\left( \varphi^{(\alpha)} \right)
	\label{eq:GLQ}
\end{align}
where $\varphi^{(\alpha)}$ and $w_\alpha$ are the $\alpha$-th node of the $K$-th polynomial and the associated weight, respectively.
In this work, we use the Gauss-Legendre quadrature for discretization.
The discretized local tensor can be expressed as
\begin{align}
	T_{ijkl} =\frac{\sqrt{w_i w_j w_k w_l}}{(2\pi)^2}
	\mathcal T\left( \varphi^{(i)},\varphi^{(j)},\varphi^{(k)},\varphi^{(l)} \right),
\end{align}
and we get a finite dimensional tensor network
\begin{align}
	Z \approx \sum_{\{\alpha\}}\prod_x T_{\alpha_{x,1}\alpha_{x+\hat 1,2}\alpha_{x+\hat 2}\alpha_{x,2}},
\end{align}
where $\{\alpha\}$ represents a set of indices associated with the Gauss-Legendre quadrature\footnote{Application of the plain Gauss-Legendre quadrature method to this model was originally proposed by Yuya Shimizu.}. 

\subsection{Improved method}
We have developed further improvement for the above method.
In the singular value decomposition (SVD) procedure to prepare the initial tensor before starting the iterative TRG steps \cite{n1-wz,phi4_quad,cphi4},
we employ the following eigenvalue decomposition:
%\begin{widetext}
\begin{align}
	M_{ijkl}=\frac{\sqrt{w_i w_j w_k w_l}}{(2\pi)^4}\int_{-\pi}^\pi d\varphi_1 d\varphi_2
	\mathcal T \left( \varphi^{(i)},\varphi^{(j)},\varphi_1,\varphi_2 \right)
	\mathcal T^\ast \left( \varphi^{(k)},\varphi^{(l)},\varphi_1,\varphi_2 \right),
	\label{eq:EVD}
\end{align}
%\end{widetext}
which is essentially equivalent to 
\begin{align}
	M_{ijkl} =\lim_{K^\prime\rightarrow\infty} \sum_{m,n=1}^{K^\prime} T_{ijmn}T^\ast_{klmn}.
\end{align}
This procedure is expected to reduce the discretization errors in $M_{ijkl}$.

To evaluate Eq.~(\ref{eq:EVD}), we use the character expansion~\cite{theta_ce1,theta_ce2}:
\begin{align}
	\mathcal T(\varphi_1,\varphi_2,\varphi_3,\varphi_4)
	=\sum_{m,n=-\infty}^\infty e^{in(\varphi_1+\varphi_2-\varphi_3-\varphi_4)}
	I_m(\beta) J_{n-m}(\theta)
\end{align}
where $I_m(\beta)$ is the $m$-th order modified Bessel function of the first kind and
\begin{align}
	J_n(\theta) =(-1)^n \frac{2}{\theta+2\pi n} \sin\left(\frac{\theta}{2}\right).
\end{align}
Then, Eq.~(\ref{eq:EVD}) is rewritten as
%\begin{widetext}
\begin{align}
	M_{ijkl}
	=\frac{\sqrt{w_i w_j w_k w_l}}{(2\pi)^4}
	\sum_{n=-\infty}^\infty e^{in(\varphi^{(i)}+\varphi^{(j)}-\varphi^{(k)}-\varphi^{(l)})}
	\left( \sum_{m,m^\prime=-\infty}^\infty I_m(\beta)I_{m^\prime}(\beta) J_{n-m}(\theta)J_{n-m^\prime}(\theta) \right).
\end{align}
%\end{widetext}
In the practical calculation,
the sums of $n,m$ and $m^\prime$ can be truncated when the contributions of the terms are small enough. In this work we discard the contributions of $I_{m,m^\prime}/I_0 < 10^{-12}$ or $J_{n-m,n-m^\prime}/J_0 < 10^{-12}$.

\section{Numerical analysis} 
\label{sec:analysis}
 
\subsection{Setup}
\label{subsec:setup}

The partition function of Eq.~(\ref{eq:PF}) is evaluated with the TRG method at $\beta=$0.0 and 10.0 as a function of $\theta$ on a $V=L\times L$ lattice, where $L$ is enlarged up to 1024.
We choose $K=32$ for the polynomial order of the Gauss-Legendre quadrature in Eq.~(\ref{eq:GLQ}).
The SVD procedure in the TRG method is truncated with $D=32$.
We have checked that these choices of $D$ and $K$ provide us sufficiently converged results for all the parameter sets employed in this work.
Since the scaling factor of the TRG method is $\sqrt{2}$, allowed lattice sizes for the partition function are $L=\sqrt{2},2,2\sqrt{2},\cdots,512\sqrt{2},1024$.
The periodic boundary condition is employed in both directions so that the topological charge $Q$ is quantized to be an integer. 

\subsection{Free energy}
\label{subsec:freeenergy}

\begin{figure}[h]
	\centering
    \includegraphics[width=80mm]{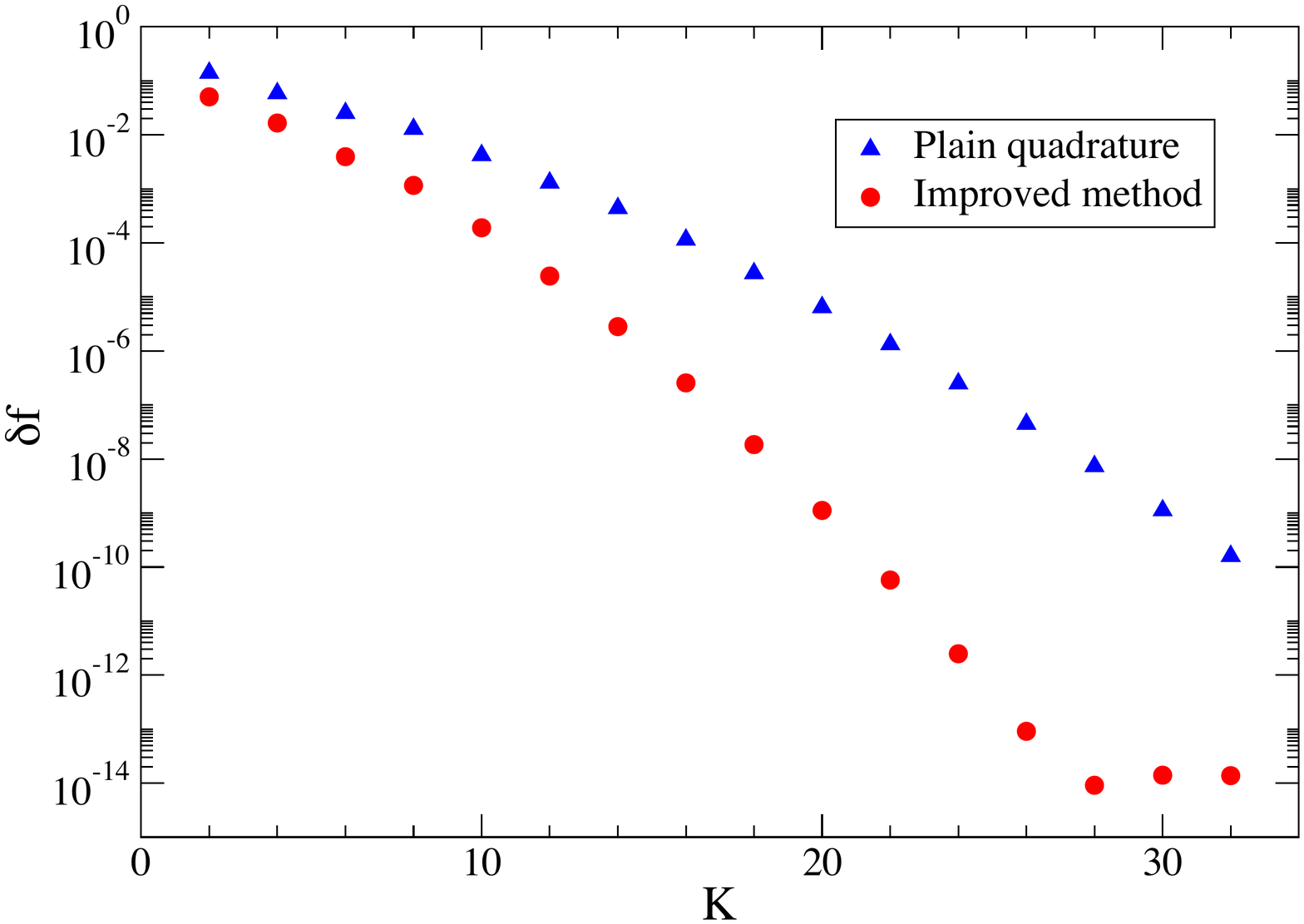}
	\caption{Relative error of free energy as a function of $K$ with $D=32$ on a $1024\times 1024$ lattice. $K$ is the polynomial order of the Gauss-Legendre quadrature in Eq.~(\ref{eq:GLQ}).}
	\label{fig:freeenergy}
\end{figure}

The analytic result for the partition function of Eq.~(\ref{eq:PF}) is given by \cite{u1_anal}:
\ben
&&Z_{\rm analytic}=\sum_{Q=-\infty}^{\infty} \left(z_{\rm P}(\theta+2\pi Q,\beta)\right)^V,\\
&&z_{\rm P}(\theta,\beta)=\int_{-\pi}^{\pi}\frac{d \varphi_{\rm P}}{2\pi}\exp\left(\beta\cos \varphi_{\rm P} +i\frac{\theta}{2\pi}\varphi_{\rm P}\right),
\een
where $z_{\rm P}(\theta,\beta)$ denotes the one-plaquette partition function with $\varphi_{\rm P}\in [-\pi,\pi]$.
In Fig.~\ref{fig:freeenergy} we plot the magnitude of the relative error for the free energy defined by
\ben
&&\delta f=\frac{\vert \ln Z_{\rm analytic} -\ln Z (K,D=32)\vert}{\vert \ln Z_{\rm analytic}\vert}
\een 
at $\theta=\pi$ on a $1024\times 1024$ lattice. There are a couple of important points to be noted. Firstly, the deviation quickly diminishes as $K$ increases even at $\theta=\pi$, around which the Monte Carlo approaches do not work effectively due to large statistical errors \cite{u1cp3-theta}. Secondly, our method yields more precise results than the plain Gauss-Legendre quadrature method at any value of $K$. Thirdly, our choice of a parameter set of $(D,K)=(32,32)$ yields $\delta f < 10^{-12}$, which means that the free energy is determined at sufficiently high precision. Hereafter we present the results obtained with $(D,K)=(32,32)$.

\subsection{Topological charge density}
\label{subsec:tc-density}

\begin{figure}[h]
    \centering
	\includegraphics[width=80mm]{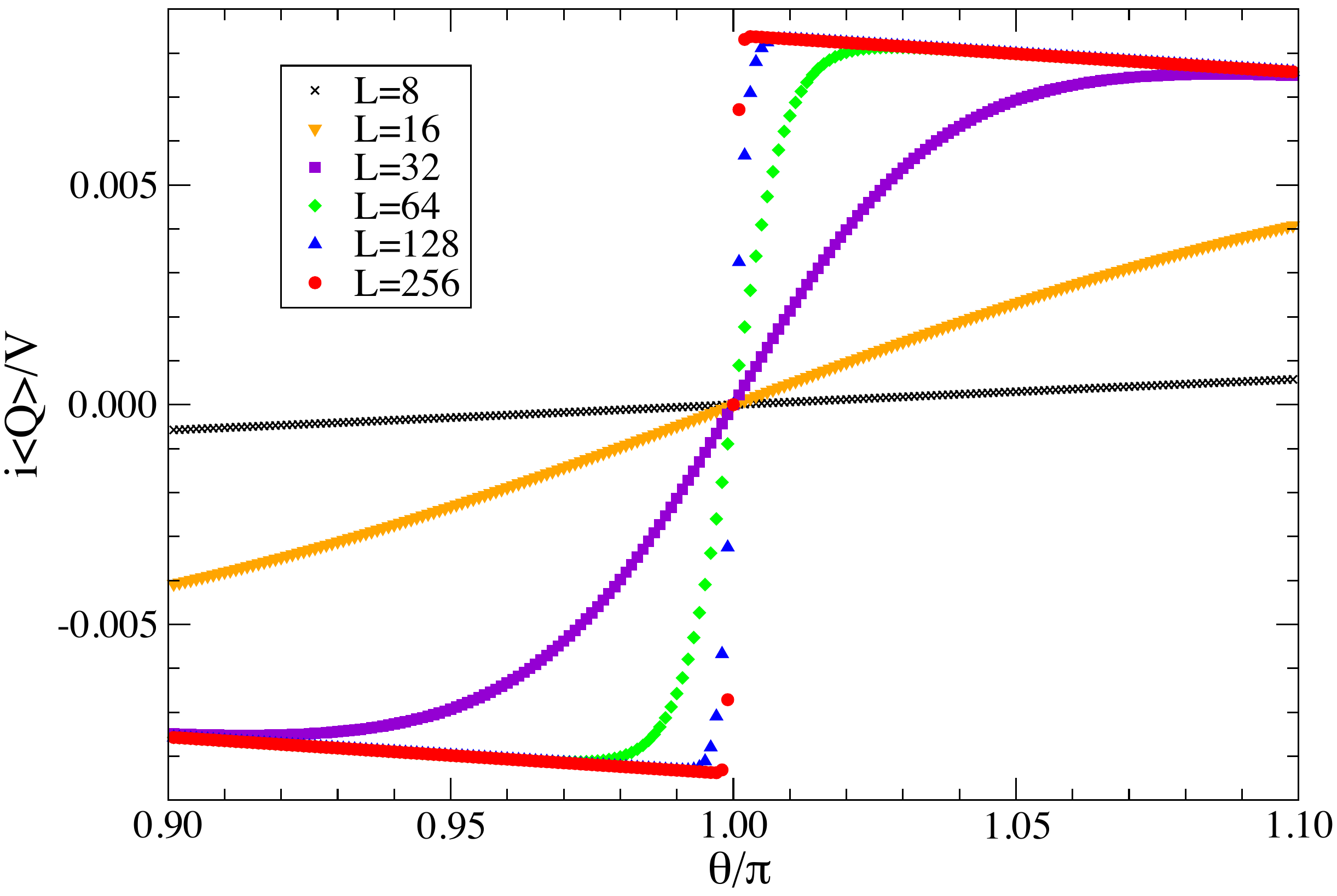}
    \caption{Topological charge density with $8\le L\le 256$ as a function of $\theta$ at $\beta=10.0$.}
    \label{fig:tc}
  \end{figure}

The expectation value of the topological charge $\la Q\ra$ at $\beta=10.0$ is obtained by the numerical derivative of the free energy with respect to $\theta$:
\ben
&&\la Q\ra=-i \frac{\partial \ln Z}{\partial \theta}.
\een 
In Fig.~\ref{fig:tc} we show the volume dependence of $\la Q\ra/V$ around $\theta=\pi$, where the analytic calculation predicts the first order phase transition at any value of $\beta$ \cite{u1_anal}. We observe that a finite discontinuity emerges with mutual crossings of curves between different volumes at $\theta=\pi$ as the lattice size $L$ is increased. This feature indicates there is a first order phase transition  at $\theta=\pi$.

\begin{figure}[h]
    \centering
	\includegraphics[width=80mm]{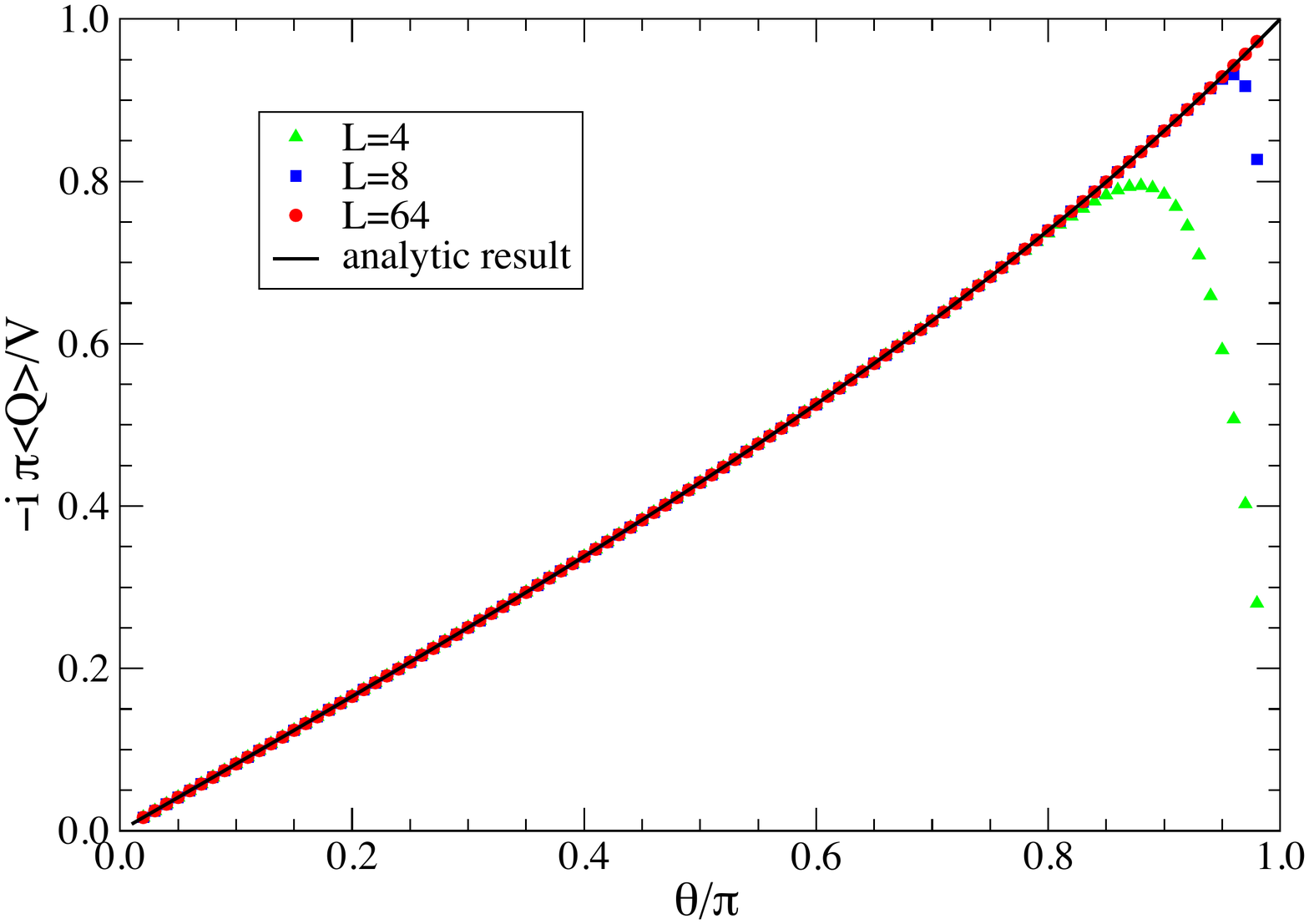}
    \caption{Topological charge density with $4\le L\le 64$ as a function of $\theta$ at $\beta=0.0$. Solid curve denotes the analytic result of Eq.~(\ref{eq:tc_b0}) obtained in the infinite volume limit.}
    \label{fig:tc_b0}
  \end{figure}

It may be interesting to calculate the topological charge density in the strong coupling limit $\beta=0.0$, whose analytical result was obtained by Seiberg in the infinite volume limit \cite{cpn_b0}:
\ben
\left. \frac{\la Q\ra}{V}\right\vert_{\beta=0}=-i\left(\frac{1}{2}\cot \left( \frac{\theta}{2}\right)-\frac{1}{\theta}\right).
\label{eq:tc_b0}
\een
Figure~\ref{fig:tc_b0} compares the numerical result at $\beta=0.0$ with the analytic expression of Eq.~(\ref{eq:tc_b0}). The discrepancy found around $\theta=\pi$ with small lattice size of $L=4$ essentially vanishes once we increase the lattice size up to $L=64$.

\subsection{Topological susceptibility}
\label{subsec:tc-suscept}

\begin{figure}[h]
    \centering
	\includegraphics[width=80mm]{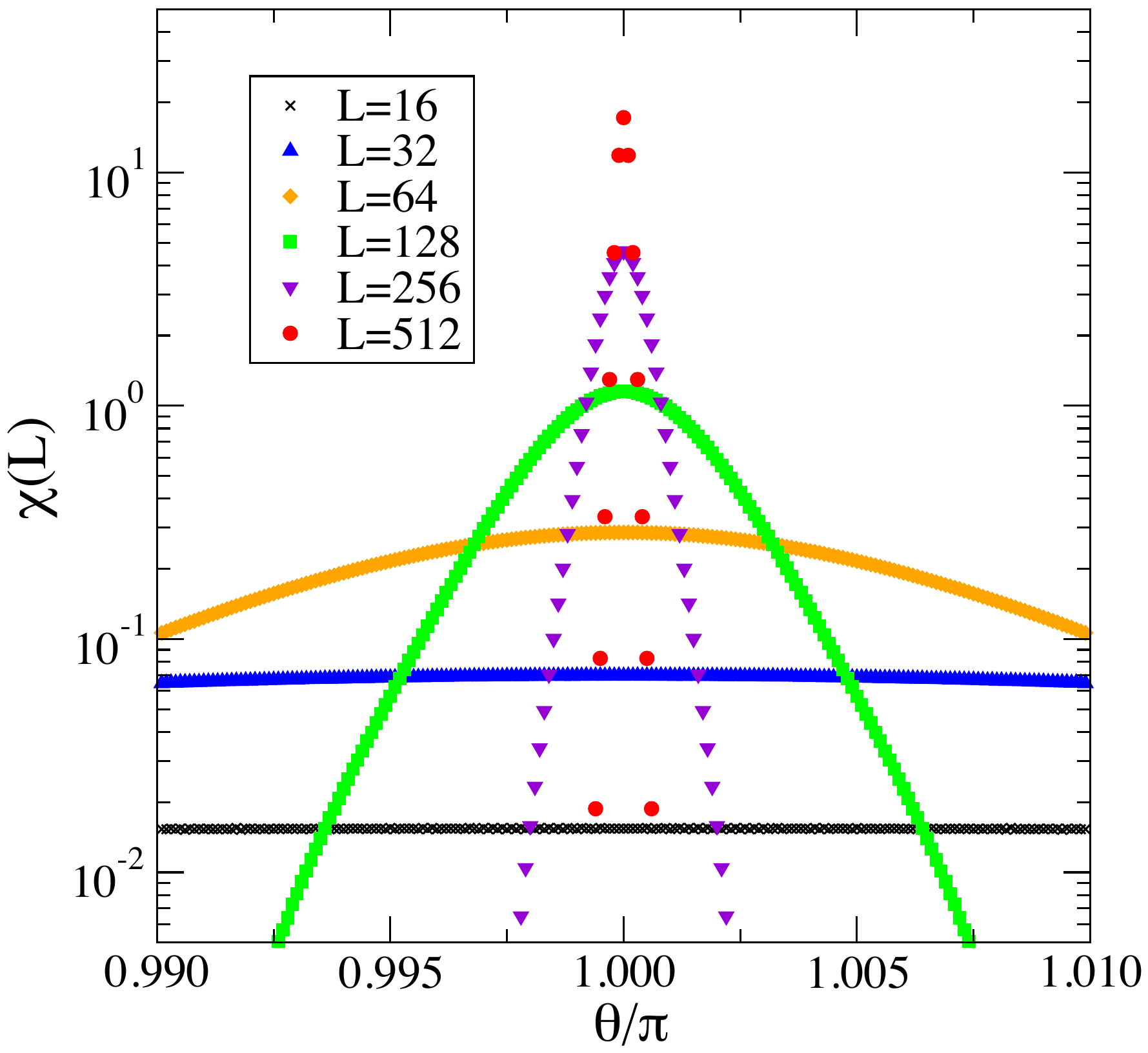}
    \caption{Topological susceptibility $\chi(L)$ as a function of $\theta$ with $16\le L\le 512$.}
    \label{fig:chi}
  \end{figure}

We investigate the properties of the phase transition by applying the finite size scaling analysis to the topological susceptibility: 
\ben
&&\chi(L)=-\frac{1}{V} \frac{\partial^2 \ln Z}{\partial \theta^2}.
\label{eq:chi}
\een 
Figure~\ref{fig:chi} shows the topological susceptibility  as a function of $\theta$ for various lattice sizes. The
peak structure is observed and its height $\chi_{\rm max}(L)$ grows as $L$ increases.
In order to determine the peak position $\theta_c(L)$ and the peak height $\chi_{\rm max}(L)$ at each $L$, we employ the quadratic approximation of the 
topological susceptibility around the peak position:
\ben
&&\chi(L)\sim \chi_{\rm max}(L)+R\left(\theta-\theta_c(L)\right)^2
\label{eq:chi_fit}
\een
with R a constant.

\begin{figure}[h]
    \centering
    \includegraphics[width=80mm]{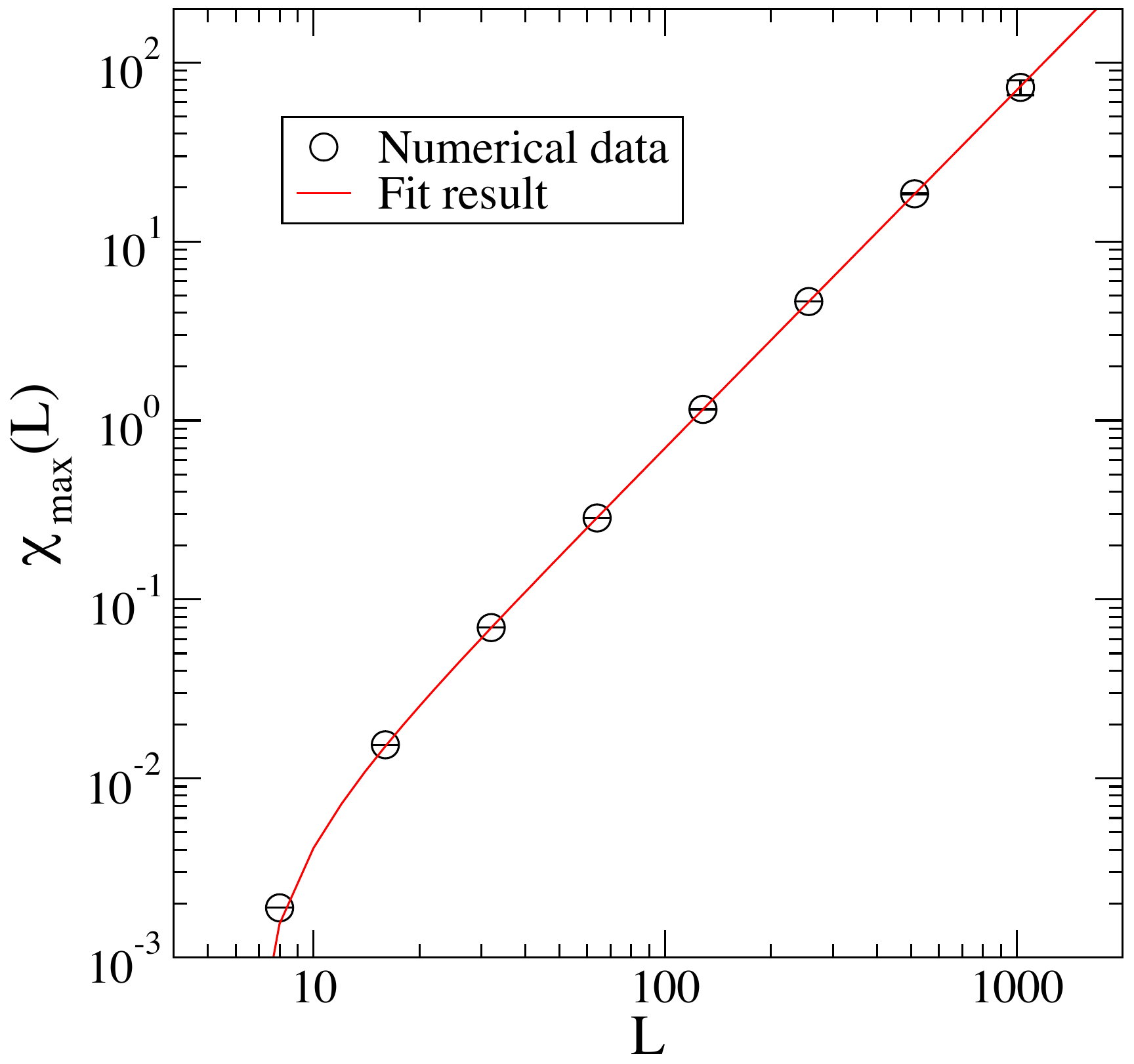}
    \caption{Peak height of topological susceptibility $\chi_{\rm max}(L)$ as a function of $L$. Solid curve denotes the fit result.}
    \label{fig:chi_max_L}
  \end{figure}

We expect that the peak height scales with $L$ as
\ben
\chi_{\rm max}(L) \propto L^{\gamma/\nu},
\een
where $\gamma$ and $\nu$ are the critical exponents. The $L$ dependence of the peak height $\chi_{\rm max}(L)$ is plotted in Fig.~\ref{fig:chi_max_L}. The solid curve represents the fit result obtained with the fit function of $\chi_{\rm max}(L) = A+ B L^{\gamma/\nu}$ choosing the fit range of $128\le L\le 1024$. The results for the fit parameters are given by $A=-3(2)\times 10^{-3},B=7.12(8)\times 10^{-5}$ and $\gamma/\nu=1.998(2)$. The value of the exponent $\gamma/\nu=1.998(2)$ is consistent with two, which is the expected critical exponent in the first-order phase transition in the two-dimensional system.

\section{Summary and outlook} 
\label{sec:summary}

We have applied the TRG method to study the 2D pure U(1) gauge theory with a $\theta$ term. The continuous degrees of freedom are discretized with the Gauss quadrature method.  We have confirmed that this model has a first-order phase transition at $\theta=\pi$ as predicted from the analytical calculation. The successful analysis of the model demonstrates an effectiveness of the Gauss quadrature approach to the gauge theories. It should be interesting to apply the TRG-based methods with the Gauss quadrature to higher dimensional gauge theories with $\theta$ term which have been hardly investigated by the Monte Carlo approach because of the sign problem. Another interesting research direction is to include fermionic degrees of freedom following the Grassmann TRG method developed in Ref.~\cite{schwinger}. This is a necessary ingredient toward investigation of the phase structure of QCD at finite density.

\begin{acknowledgments}
One of the authors (YK) thanks Yuya Shimizu for providing the results obtained by the plain Gauss-Legendre quadrature method. 
%We would like to thank ??? for providing us useful comments on the manuscript.
Numerical calculation for the present work was carried out with the Cygnus computer under the Interdisciplinary Computational Science Program of Center for Computational Sciences, University of Tsukuba.
This work is supported by the Ministry of Education, Culture, Sports, Science and Technology (MEXT) as ``Exploratory Challenge on Post-K computer (Frontiers of Basic Science: Challenging the Limits)''.
\end{acknowledgments}

%\bibliography{unsrt,apssamp}

\end{document}